\documentclass[prb,aps,twocolumn,draft,tightenlines,
showpacs,floatfix,superscriptaddress]{revtex4}
\usepackage{psfig}
\usepackage{amssymb}

\usepackage{amsmath}
\begin{document}
\title{Hole spin relaxation in semiconductor quantum dots}
\author{C. L\"u}
\affiliation{Hefei National Laboratory for Physical Sciences at
  Microscale, University of Science and Technology of China,
Hefei, Anhui, 230026, China}
\affiliation{Department of Physics, University of Science \&
Technology of China, Hefei, Anhui, 230026, China}
\altaffiliation{Mailing Address.}
\author{J. L. Cheng}
\affiliation{Department of Physics, University of Science \&
Technology of China, Hefei, Anhui, 230026, China}
\author{M. W. Wu}
\thanks{Author to whom correspondence should be addressed}
\email{mwwu@ustc.edu.cn.}
\affiliation{Hefei National Laboratory for Physical Sciences at
  Microscale, University of Science and Technology of China,
Hefei, Anhui, 230026, China}
\affiliation{Department of Physics, University of Science \&
Technology of China, Hefei, Anhui, 230026, China}
\altaffiliation{Mailing Address.}
\date{\today}

\begin{abstract}  
Hole spin relaxation time due to the hole-acoustic phonon
scattering in GaAs quantum dots confined in quantum wells
along (001) and (111) directions
is studied after the exact 
diagonalization of Luttinger Hamiltonian. 
Different effects such as strain, magnetic
field, quantum dot diameter, quantum well width and the temperature 
on the spin relaxation time are investigated thoroughly. 
Many features which are quite different from the electron spin relaxation
in quantum dots and quantum wells are presented with the underlying 
physics elaborated.

\end{abstract}

\maketitle
\section{Introduction}
Recently, considerable interests have been devoted to spin-related phenomena
in semiconductors due to the enormous potential of the 
spintronic devices.\cite{spintronics,das}  
Among these,  properties of electron spins confined
in semiconductor quantum dots (QD's) are essential to the 
proposed qubits in quantum computers and  have therefore caused much
attention.\cite{manuel,Gupta,Khaetsii,Hackens,Alexander1,Alexander}
Many works calculated the spin relaxation time (SRT) of electrons
due to the spin-orbit coupling induced spin-flip electron-phonon scattering
at very low temperatures,\cite{Woods,Khaetsii,Alexander,Alexander1} where the
dominant electron-phonon scattering
arises from the piezoelectric potential.
These works are based on perturbation theory
where the spin-orbit coupling is treated as a perturbation 
in the Hilbert space spanned by $H_0$ which does not include the spin-orbit 
coupling. Moreover only the lowest few energy levels of $H_0$ are included 
in the theory. Recently  we have shown 
that the perturbation method is inadequate in accounting for the
electron structure and therefore the SRT in semiconductor QD's:
The SRT obtained from the perturbation
approach used in the literature\cite{Woods,Khaetsii,Alexander,Alexander1} 
is several orders of magnitude smaller than the exact
value.\cite{Cheng} Very recently Wood {\em et al.} 
investigated the SRT of a hole in QD's.\cite{Woods2}  Again perturbation
method is used and only the SRT induced by the electron-phonon
scattering due to the deformation potential is considered.
Many other effects  such as strain, quantum well orientation
 and multi-subband effects
as well as the effect from the electron-phonon scattering due to 
the piezoelectric coupling, have not been studied in their work.

In the present paper, we investigate the hole SRT of
GaAs QD's confined in quantum wells along (001) and (111)
directions by parabolic potentials with strain included
by exactly diagonalizing the hole Hamiltonian.  
We calculate the hole SRT due to the scattering with  acoustic phonons
by the Fermi golden rule after 
getting the hole energy spectra and the wavefunctions from the 
exact diagonalization. We discuss how the strain, QD radius,
magnetic field, temperature and quantum well width affect the SRT. 
We show that strain on quantum wells of different growth directions
affects the QD spin relaxation in totally different ways: 
for QD's in (001) quantum well strain changes the relative position of energy levels of heavy hole
and light hole; but for those in (111) quantum well, 
strain makes additional spin mixing and induces additional spin relaxation.
Also we show that unlike the case of electrons where the SRT
is mainly determined by the electron-phonon scattering due to the
piezoelectric interactions, for holes  both the hole-phonon coupling
due to piezoelectric interaction and that due to  deformation  potential
make important contributions to the spin relaxation process, although
their relative importance  changes under different conditions.

The paper is organized as follows: In Sec.\ II we set up our model and
Hamiltonian. In Sec.\ III we present our numerical
results.  We discuss the SRT of QD's in (001) quantum well in
 Sec.\ III(A). We first discuss a simple case: a small QD without
strain where we compare our results with those obtained from 
the perturbation method. Then we discuss strain dependence of the SRT
when the confinement of the quantum well is very strong and there is
only one subband. We finish Sec.\ III(A) by showing the strain, magnetic
field, and QD radius dependence of the SRT in the case of
large well width  (multi-subband effects).
Then we turn to the case of QD's in (111) 
quantum well in Sec.\ III(B).  In
Sec.\ III(C) we show the well 
width dependence of the SRT in both (001) and (111) quantum wells. 
We conclude  in Sec.\ IV.

\section{Model and Hamiltonian}

We use a simplified model to study the spin relaxation in  QD's
which are confined by parabolic potentials $V_c({\bf
  r})$ in a quantum well of
width $a$. Due to the confinement of the quantum well, 
the momentum states along $z$ axis are quantized. 
With the hard-wall approximation, the hole momentum states along $z$-axis are
therefore characterized by the subband index $n_z$. 
 The total Hamiltonian is given by
\begin{equation}
  \label{total_Hamiltonian}
  H = H_h+H_{strain}+H_{ph}+H_{int}\ ,
\end{equation}
in which $H_h$ is the 4$\times$4  Luttinger Hamiltonian for holes.
When the growth direction of the quantum well is along the 
(001) direction ($z$-axis) and  the matrix elements are 
arranged in the order of $J_z =
+\frac{3}{2}$, $+\frac{1}{2}$, $-\frac{1}{2}$ and $-\frac{3}{2}$,
$H_h$ can be written as\cite{Trebin}
\begin{widetext}
\begin{equation}
  \label{Luttinger_Hamiltonian}
  H_{h} = \frac {1} {2m_0} \left( \begin{array}{cccc} P + Q + 3 \hbar e
  B \kappa & S & R &
  0 \\ S^\dagger & P - Q - \hbar e B \kappa & 0 & R \\ R^\dagger & 0 &
  P - Q + \hbar e B \kappa & -S \\ 0 & R^\dagger & -S^\dagger & P + Q - 3
  \hbar e B \kappa  \end{array} \right)+V_c({\bf r})\ ,
\end{equation}
in which
\begin{equation}
  \label{potential}
  V_c({\bf r}) = \left( \begin{array}{cccc} \frac{1}{2} m_{h\|}^{001}
  (\omega_h^{001})^2 r^2 & 0 & 0 & 0 \\ 0 & \frac{1}{2} m_{l\|}^{001}
  (\omega_l^{001})^2 r^2 & 0 & 
  0 \\ 0 & 0 & \frac{1}{2} m_{l\|}^{001} (\omega_l^{001})^2 r^2& 0 
\\ 0 & 0 & 0 &
  \frac{1}{2} m_{h\|}^{001} (\omega_h^{001})^2 r^2
  \end{array} \right)\ ,
\end{equation}
\end{widetext}
and
\begin{eqnarray}
  \label{PQ}
&&  P\pm Q = (\gamma_1 \pm \gamma_2)[P_x^2
  + P_y^2]
 \nonumber \\ &&
 \mbox{}\hspace{1cm}
+(\gamma_1 \mp 2 \gamma_2) \frac{\hbar^2 \pi^2 n_z^2}{
    a^2} \delta_{n_z,n_z^{\prime}}\ ,\\
  \label{S}
&&  S = -  2 \sqrt{3} \gamma_3 \frac{4i\hbar n_z^{\prime}n_z}{a
    [(n_z^{\prime})^2-(n_z)^2]}(1-\delta_{n_z,n_z^{\prime}})\nonumber \\
&&\mbox{}\hspace{1cm}\times [P_x -
  iP_y]\ ,\\
  \label{R}
&&  R = - \sqrt{3} \{\gamma_2 [P_x^2 - P_y^2]   \mbox{}- 2 i \gamma_3
  P_xP_y \}\ .
\end{eqnarray}
In these equations, $m_0$  denotes free electron mass;
 $\gamma_1$, $\gamma_2$, $\gamma_3$ and $\kappa$ are Luttinger
coefficients; and $n_z$ and $n_z^\prime$ represent the subband indices.
$\omega_h^{001}$ and $\omega_l^{001}$ in the two dimensional confinement 
potential $V_c({\bf r})$ [Eq.\ (\ref{potential})] 
represent the confinements experienced by
the heavy hole and light hole respectively and 
are given by $\omega_h^{001}
= \hbar/(m_{h\|}^{001} d^2)$ and $\omega_l^{001}
=\hbar/(m_{l\|}^{001} d^2)$, with
 $m_{h\|}^{001}=m_0/(\gamma_1+\gamma_2)$ and 
$m_{l\|}^{001}=m_0/(\gamma_1-\gamma_2)$ 
standing for the effective masses of heavy hole and light hole in 
the direction  perpendicular to the
growth (001) direction and $d$
representing the QD diameter.
By applying a magnetic field ${\bf B}$ along the growth ($z$)
direction of the quantum well  and adopting  the Coulomb  gauge 
${\bf A}=(-\frac{By}{2},\frac{Bx}{2}, 0)$, one has
$P_x=(\hbar k_x+\frac{eBy}{2})$ and $P_y=(\hbar
 k_y-\frac{eBx}{2})$. 

From the Luttinger Hamiltonian Eq.\ (\ref{Luttinger_Hamiltonian})
one can see that when the well width $a$ is sufficiently small and
only the lowest subband in QD is important, $S=0$ and 
$+\frac{3}{2}$ ($-\frac{3}{2}$) states can only mix with 
$-\frac{1}{2}$ ($+\frac{1}{2}$) states.
Therefore there is no mixing between the spin-up and
-down  ($\pm \frac{3}{2}$) heavy hole  states
and between the spin-up and -down ($\pm \frac{1}{2}$) light hole states. 
Spin mixing between the spin-up heavy hole and the spin-down light hole
or {\em vice versa}
is negligible as the energy
difference between the heavy hole and  light hole states is
usually too large.
Nevertheless  for larger well width where higher subbands 
are needed, $S$ no longer equals to zero and the
spin-up and -down heavy-hole states are mixed with each other,
mediated by the light-hole states. The same is true for the 
spin-up and -down light-hole states.

The strain Hamiltonian $H_{strain}$ given by the Bir-Pikus 
Hamiltonian\cite{Bir} has the form:
\begin{equation}
  \label{strain_Hamiltonian}
  H_{strain} = \left( \begin{array}{cccc} F & H & I & 0 \\ H^\dagger & G & 0 &
  I \\ I^\dagger & 0 & G & -H \\ 0 & I^\dagger & -H^\dagger & F
  \end{array} \right)\ 
\end{equation}
with the matrix elements being
\begin{eqnarray}
  \label{F}
  F&=&-(D_a+\frac{D_b}{2}) \mbox{Tr}( \epsilon ) + \frac{3D_b}{2} 
\epsilon_{zz}\ ,\\
  \label{G}
  G&=&-(D_a-\frac{D_b}{2}) \mbox{Tr}( \epsilon ) - \frac{3D_b}{2} 
\epsilon_{zz}\ ,
\\
  \label{H}
  H&=&D_d(\epsilon_{zx} - i \epsilon_{zy})\ ,\\
  \label{I}
  I& =& \frac{\sqrt{3}}{2} D_b (\epsilon_{xx} - \epsilon_{yy} ) - i D_d
  \epsilon_{xy}\ .
\end{eqnarray}
Here $D_a$, $D_b$ and $D_d$ are the deformation potential constants.
$\epsilon$ is the strain tensor with  $\epsilon_{ij}$ denoting
the tensor components. 
For (001)-oriented zinc blende crystal the strain tensor
components are given by\cite{Park}
\begin{eqnarray}
  \label{001}
&&  \epsilon_{xx}^{001}= \epsilon_{yy}^{001}=
  \epsilon_{\|}=\frac{a_2 - a_1}{a_1} \\
\label{001z}
&& \epsilon_{zz}^{001} = -2
  \frac {C_{12}}{C_{11}} \epsilon_{\|} \\ 
&& \epsilon_{xy}^{001} =
  \epsilon_{yz}^{001} = \epsilon_{zx}^{001} =0 \ ,
\end{eqnarray}
where $a_1$ and $a_2$ are the lattice constants of epilayer (GaAs)
and substrate materials, 
and $C_{11}$ and $C_{12}$  are the stiffness
constants. One can
see that for (001)-oriented zinc blende crystal, $H=I=0$ and
the strain Hamiltonian Eq.\ (\ref{strain_Hamiltonian}) has only the
diagonal terms. Therefore the strain
does not induce any  extra spin mixing, but adjusts the relative positions of
the heavy-hole and light-hole energy levels. 

When the growth direction of the quantum well ($z$-axis) is along 
(111) direction,
the hole Luttinger Hamiltonian $H_h$ is the same as
that in Eq.\ (\ref{Luttinger_Hamiltonian}), but with the matrix
elements being replaced by\cite{Trebin}
\begin{widetext}
\begin{eqnarray}
  \label{111PQ}
&&P \pm Q = (\gamma_1 \pm \gamma_3)[P_x^2
  + P_y^2]
+(\gamma_1 \mp 2 \gamma_3) \frac{\hbar^2 \pi^2 n_z^2}{
    a^2} \delta_{n_z,n_z^{\prime}}\ ,\\
  \label{111S}
&& S=\frac{\sqrt{6}}{3}(\gamma_2-\gamma_3)[P_x +
    iP_y]^2
 -  \frac{2 \sqrt{3}}{3}
    (2\gamma_2+\gamma_3) \frac{4i\hbar 
    n_z^{\prime}n_z}{a
    [(n_z^{\prime})^2-(n_z)^2]}(1-\delta_{n_z,n_z^{\prime}})
[P_x -
  iP_y]\ ,\\ 
  \label{111R}
&&  R= - \frac{\sqrt{3}}{3} (\gamma_2+2 \gamma_3) [P_x -
  iP_y]^2 + \frac{2\sqrt{6}}{3}
  [(\gamma_2-\gamma_3) 
  \frac{4i\hbar
    n_z^{\prime}n_z}{a [(n_z^{\prime})^2-(n_z)^2]}(1-\delta_{n_z,n_z^{\prime}})
[P_x +  iP_y]]\ .
\end{eqnarray}
\end{widetext}
Moreover $\omega_h^{001}$, $\omega_l^{001}$, $m_{l\|}^{001}$ and
$m_{h\|}^{001}$ in Eq.\ (\ref{potential}) should be replaced by
$\omega_h^{111}$, $\omega_l^{111}$, $m_{l\|}^{111}$ and
$m_{h\|}^{111}$, which are given by  $\omega_h^{111}
= \hbar/({m_{h\|}^{111} d^2})$ and $\omega_l^{111}
= \hbar/({m_{l\|}^{111} d^2})$,
 $m_{h\|}^{111}=m_0/(\gamma_1+\gamma_3)$ and 
$m_{l\|}^{111}=m_0/(\gamma_1-\gamma_3)$ respectively. 
From Eqs.\ (\ref{111PQ}-\ref{111R}), one finds that differing from the
previous (001) case, here $S$ and $R$ are
nonzero even for the single subband case.
This  means when the
growth direction is along (111) crystal direction, 
there is always mixing between the
spin-up  and -down heavy-hole states and the spin-up and -down
light-hole states.

The strain Hamiltonian $H_{strain}$ for (111)-oriented zinc blende
crystal is same as that in Eqs.\ (\ref{strain_Hamiltonian}-\ref{I}), 
but now the strain tensor components are given by\cite{Park}
\begin{eqnarray}
  \label{111}
&& \epsilon_{xx}^{111} =
  \epsilon_{yy}^{111} = \epsilon_{zz}^{111} =\frac{1}{3}
  (2-1/\sigma^{111}) \epsilon_{\|}\ ,
\\ 
&& \epsilon_{xy}^{111} =
  \epsilon_{yz}^{111} = \epsilon_{zx}^{111} =-\frac{1}{3}
  (1+1/\sigma^{111}) \epsilon_{\|}\ ,
\\ 
\label{s111}
&& \sigma^{111}=\frac{C_{11}+3C_{12}+4C_{44}}{2C_{11}+4C_{12}-4C_{44}}\ ,
\end{eqnarray}
with $C_{44}$ the stiffness constant.
Substituting Eqs.\ (\ref{111}-\ref{s111}) into 
Eqs.\ (\ref{strain_Hamiltonian}-\ref{I}), one gets $F=G$ and
the off-diagonal elements $H$ and $I$ are no longer equal to zero. Consequently
when one applies a strain on (111)-oriented zinc blende quantum well, the
change of energy levels of the light and heavy holes  are 
almost the same and the strain introduces additional spin mixing
between the spin-up and -down heavy/light holes.

$H_{ph}$ in Eq.\ (\ref{total_Hamiltonian}) is
the Hamiltonian of acoustic phonons and is given by $H_{ph}= \sum_{{\bf 
 Q}\lambda}\hbar\omega_{{\bf Q}\lambda}a^{\dagger}_{{\bf
 Q}\lambda}a_{{\bf Q}\lambda}$ with $\omega_{{\bf Q}\lambda}$ 
standing for the phonon energy spectrum
of branch $\lambda$ and momentum ${\bf Q}$. 
Two different hole-phonon scattering mechanisms contribute 
to the spin relaxation
for the temperatures we consider here. One is
hole-phonon scattering due to piezoelectric coupling which is given by
\begin{equation}
\label{ep}
H_{int}^{pie} = \sum_{{\bf Q}\lambda}M_{{\bf
 Q}\lambda}(a^{\dagger}_{-{\bf Q}\lambda}+a_{{\bf Q}\lambda})\exp(i{\bf
Q}\cdot{\bf r}) \,
\end{equation}
with $M_{{\bf Q}\lambda}$ being the scattering the matrix elements.
For longitudinal acoustic phonons  $M^2_{{\bf Q}pl} =
 \frac{32\pi^2e^2e^2_{14}}{\varepsilon^2\rho v_{sl}}
 \frac{(3Q_xQ_yQ_z)^2}{Q^7}$ and for  two
transverse acoustic phonons   $\sum_{j=1,2}M^2_{{\bf Q}pt_j} =
 \frac{32\pi^2e^2e^2_{14}}{\varepsilon^2\rho v_{st}Q^5}
[Q_x^2Q_y^2+Q_y^2Q_z^2+Q_z^2Q_x^2-\frac{(3Q_xQ_yQ_z)^2}{Q^2}]$.
Here $\rho$ is the GaAs volume density, $e_{14}$ represents the
piezoelectric coupling constant and $\varepsilon$ denotes the static dielectric
 constant. The acoustic phonon energy spectra $\omega_{{\bf Q}\lambda}$ are 
given by $\omega_{{\bf Q}l}= v_{sl}Q$ for the longitudinal mode 
and $\omega_{{\bf Q}pt} = v_{st}Q$ for the transverse mode with 
$v_{sl}$ and $v_{st}$ standing for the corresponding sound
velocities. $Q=\sqrt{Q_x^2+Q_y^2+Q_z^2}$.
It is noted that this kind of scattering {\em does not} flip hole spin 
and therefore only when the hole wave function itself contains spin mixing
can $H_{int}^{pie}$ contributes to spin relaxation. 
The other is hole-phonon scattering due to 
the deformation  potential $H_{int}^{def}$.
$H_{int}^{def}$ can be derived 
from $H_{strain}$  Eq.\ (\ref {strain_Hamiltonian}) by substituting 
$\epsilon_{ij}$ in Eqs.\ (\ref{F}-\ref{I}) 
by $\epsilon_{ij}=\epsilon_{ij}^0
+\epsilon_{ij}^\prime$, and splitting $H_{strain}$ into two
parts: one contains all terms proportional to $\epsilon_{ij}^0$
and the other contains all terms proportional to $\epsilon_{ij}^\prime$.\cite{Woods2}
The second part is therefore $H_{int}^{def}$
if $\epsilon_{ij}^0$ represent the
strain tensor components caused by the sample and
$\epsilon_{ij}^{\prime}$ are the tensor components caused by the lattice vibrations.
$\epsilon_{ij}^{\prime}$ can further be written in terms of normal-mode
coordinates as  
\begin{eqnarray}
  \label{epsilon_ij}
  \epsilon_{ij}^{\prime} &=& \sum_{{\bf Q}\lambda} 
\frac{i}{2} \sqrt{ \frac{\hbar} {2\rho
      V \omega_{{\bf Q}\lambda}}} ( a_{{\bf Q}\lambda} + a_{ - {\bf
      Q}\lambda}^{\dagger}) \nonumber \\ && \times ( \hat{\eta}_{i\lambda}
 Q_j +  \hat{\eta}_{j\lambda}
      Q_i)e^{i{\bf Q \cdot r}} \ ,
\end{eqnarray}
with $\rho$ being the mass density of the material, ${\bf Q}$ standing
for the phonon wave vector and $\hat{\eta}_{i\lambda}$ 
representing the unit vector of polarization
of $\lambda$-phonon  along $i$-direction. 
For longitudinal mode  $\hat{\eta_i}=Q_i/Q$ and for two transverse modes
$\hat{\eta_{t1}}=(Q_xQ_z,Q_yQ_z,-Q_{\bot}^2)/QQ_{\bot}$ and
$\hat{\eta_{t2}}=(Q_y,-Qx,0)/Q_{\bot}$ with
 $Q_{\bot}=\sqrt{Q_x^2+Q_y^2}$.\cite{Woods2} 
It is pointed out  that $H_{int}^{def}$  is spin-flip scattering as
it contains non-zero off-diagonal parts $H$ and $I$, 
and therefore can cause spin relaxation
even for the case when there is no spin mixing
in the wavefunctions. Accounting for these two hole-phonon
scattering mechanisms, one has $H_{int}=
H_{int}^{pie}+H_{int}^{def}$.

We diagonalize the Luttinger Hamiltonian $H_h$ in the Hilbert 
space  $|n,l,n_z,\sigma \rangle$  constructed by 
$H_0$ which is taken to be the diagonal part of  $H_h$:
\begin{equation}
\label{psiell}
|\Psi_\ell\rangle=\sum_{nln_z\sigma}C_{nln_z\sigma}^\ell
|n,l,n_z,\sigma\rangle\ .
\end{equation}
Here
$H_0|n,l,n_z,\sigma\rangle=E_{n,l,n_z,\sigma}^\xi|n,l,n_z,\sigma\rangle$
with
\begin{eqnarray}
\label{eigen}
\langle {\bf r}|n,l,n_z,\sigma\rangle&=&N_{n, l}(\sqrt{\alpha}
   r)^{|l|}e^{-\frac{\alpha r^2}{2}}L^{|l|}_n(\alpha
   r^2)e^{il\theta}\nonumber \\ && \times \sqrt{\frac{2}{a}}
\sin(\frac{n_z\pi}{a}z)\ ,\\
  \label{energy_heavyhole}
  E_{n,l,n_z,\pm \frac{3}{2}}^{\xi} &=& \frac{m_0}{m_{h\|}^{\xi}}
  [\hbar\Omega(2n + |l|
  + 1)-\hbar\omega_B l] \nonumber \\ &&\mbox{}\pm  \frac{3\hbar e B
  \kappa} {2m_0} + 
  \frac{m_0}{m_{hz}^{\xi}} \frac{\hbar^2 \pi^2 n_z^2}{2 m_0 a^2} \ ,
\\
  \label{energy_lighthole}
  E_{n,l,n_z,\pm \frac{1}{2}}^{\xi} &=& \frac{m_0}{m_{l\|}^{\xi}}
[\hbar\Omega(2n + |l|
  + 1)-\hbar\omega_B l] \nonumber \\ 
&&\mbox{}\pm  \frac{\hbar e B \kappa}{2m_0} +
 \frac{m_0}{m_{lz}^{\xi}} \frac{\hbar^2 \pi^2 n_z^2}{2 m_0 a^2} \ .
\end{eqnarray}
In these equations $n=0, 1, 2,\cdots$ and $l=0,\pm 1,\pm 2, \cdots$
are quantum numbers; $\xi$ denotes the growth direction which can be (001)
 or (111); $m_{hz}^{\xi}$ and $m_{lz}^{\xi}$ stand for the
effective masses of heavy and light holes in the $z$-direction
which are given by
$m_0/m_{hz}^{001}=\gamma_1-2\gamma_2$,
$m_0/m_{hz}^{111}=\gamma_1-2\gamma_3$,
$m_0/m_{lz}^{001}=\gamma_1+2\gamma_2$ and 
$m_0/m_{lz}^{111}=\gamma_1+2\gamma_3$;
$\Omega=\sqrt{\omega_0^2+\omega^2_B}$ 
with $\omega_0=\hbar/(m_0d^2)$ and  $\omega_B=eB/(2m_0)$;
$N_{n,l} = \left(\frac{\alpha n!}
{\pi(n+|l|)!}\right)^{\frac{1}{2}}$ with $\alpha
 = m_0\Omega/\hbar$. 
$L_n^{|l|}$ is the generalized Laguerre polynomial.

The eigenfunction $|\Psi_\ell\rangle$ in Eq.\ (\ref{psiell})
is a mixture of four different components: spin-up and -down heavy-hole 
and light-hole states.
We assign an eigenstate $\ell$ to be spin up if the
spin-up components are larger than the spin-down ones. An
hole at initial state $i$ with energy $\epsilon_i$ and a
spin polarization can be scattered by the phonon into another
state $f$ with energy $\epsilon_f$ and the opposite spin
polarization. The rate of such scattering can be 
described by the Fermi golden rule:
 \begin{eqnarray}
   \label{fermirule}
 \Gamma_{i\to f}&=&\frac{2\pi}{\hbar}\sum_{{\bf Q}\lambda}|
{\cal M}_{{\bf Q}\lambda}|^2
[{\bar n}_{{\bf
   Q}\lambda}\delta(\epsilon_f-\epsilon_i-\omega_{{\bf Q}\lambda})\nonumber\\
 &&+
 ({\bar n}_{{\bf
   Q}\lambda} + 1)\delta(\epsilon_f-\epsilon_i+\omega_{{\bf Q}\lambda})]\ ,
 \end{eqnarray}
with ${\bar n}_{{\bf Q}\lambda}$ representing the Bose distribution of phonon
with mode $\lambda$ and momentum ${\bf Q}$ at temperature
$T$ and ${\cal M}_{{\bf Q}\lambda}$ being the corresponding matrix elements.
For hole-phonon scattering due to the
piezoelectric coupling, $|{\cal M}_{{\bf Q}\lambda}|^2=
|M_{{\bf Q}\lambda}\langle f|e^{i{\bf Q}\cdot{\bf r}}|i\rangle|^2$.
It is noted that only when the eigenstates $|i\rangle$ and $|f\rangle$
contain spin mixing can ${\cal M}_{{\bf Q}\lambda}\not=0$ be possible.
For hole-phonon scattering due to the deformation  
potential, $|{\cal M}_{{\bf 
  Q}\lambda}|^2=|\langle f|H_{{\bf Q}\lambda}|i\rangle|^2$ with 
$H_{{\bf Q}\lambda}$  the matrix 
for the hole-deformation potential. As $H_{{\bf Q}\lambda}$  itself 
contains spin mixing, therefore it is not necessary to have spin-mixed
initial and final states to ensure ${\cal M}_{{\bf Q}\lambda}\not=0$.
The total SRT $\tau$ can be written as
\begin{equation}
  \label{SRT}
  \frac{1}{\tau} = \sum_{i}f_i\sum_f\Gamma_{i\to f}\ ,
\end{equation}
in which $f_i = C\exp[-\epsilon_i/(k_BT)]$ denotes the Maxwell distribution
of the $i$-th level with $C$ being a constant.

\section{Numerical Results}

It is seen from our previous discussion that  hole spin relaxation in QD's
is very complicated and is  affected by many effects. When the quantum
well is along (001) direction, for small well width  (with only the 
lowest subband) only hole-phonon scattering due to deformation potential
contributes to the spin relaxation; for large well width (with multi subbands)
hole-phonon scattering due to both  deformation potential and piezoelectric
coupling contributes to the spin relaxation. Strain itself in this case
cannot bring additional spin relaxation but influences it by
 changing or even reversing the relative 
positions of energy levels of heavy hole and light hole. 
Especially when a minus strain makes the energy levels of 
heavy hole and light hole very close to each other, spin mixing
between spin-up (-down) heavy hole and spin-down (-up) light hole 
cannot be neglected anymore and therefore the hole-phonon scattering
due to piezoelectric coupling may contribute to spin relaxation also.
Nevertheless, when the quantum well is along (111) direction,
regardless  of the well width, there exists spin mixing 
between spin-up and -down heavy (light) holes in the eigen functions 
of the Luttinger Hamiltonian. Therefore hole-phonon scattering due 
to both  deformation potential and piezoelectric
coupling contributes to the spin relaxation. Moreover, strain itself
in this case makes additional spin mixing and induces additional 
spin relaxation. In this section we perform a comprehensive 
investigation to find out the relative importance of above
mentioned effects under various conditions such as temperature,
magnetic field, QD radius and quantum well width. 
The parameters used in the computation are 
listed in Table\ \ref{parameter}.\cite{made,oh,Bir}

In order to ensure the convergence of the energy spectra
$\epsilon_{\ell}$, we use sufficient basis functions to diagonalize
the hole Hamiltonian $H_h$. For example, in a QD with $B=1$\ T, $a=5$\ nm, 
$d=20$ nm and without strain one has to use 100 basis functions
to converge the lowest 40 levels; Nevertheless, 
when $a=20$\ nm one has to use 484 basis functions to
converge the same levels.

\begin{table}[htbp]
\begin{tabular}{lllllllllllllll}\hline\hline
$\rho$&\mbox{}&$5.3\times10^3$ kg/m$^3$&\mbox{}&\mbox{}&\mbox{}&$\epsilon$&\mbox{}&12.9&\mbox{}&\mbox{}&\mbox{}&$\gamma_1$&\mbox{}&6.85\\
$v_{st}$&\mbox{}&$2.48\times10^3$\ m/s&\mbox{}&\mbox{}&\mbox{}&$D_a$&\mbox{}
&$-6.7$ eV&\mbox{}&\mbox{}&\mbox{}&$\gamma_2$&\mbox{}&2.1\\
$v_{sl}$&\mbox{}&$5.29\times10^3$\ m/s&\mbox{}&\mbox{}&\mbox{}&$D_b$&\mbox{}
&$-1.7$ eV&\mbox{}&\mbox{}&\mbox{}&$\gamma_3$&\mbox{}&2.9\\
$e_{14}$&\mbox{}&$1.41\times10^3$\
V/m&\mbox{}&\mbox{}&\mbox{}&$D_d$&\mbox{}
&$-4.55$ eV&\mbox{}&\mbox{}&\mbox{}&$\kappa$&\mbox{}&1.2\\
$C_{11}$&\mbox{}&$11.81$&\mbox{}&\mbox{}&\mbox{}&$C_{12}$&\mbox{}&5.32&\mbox{}&\mbox{}&\mbox{}&$C_{44}$&\mbox{}&5.94\\ 
\hline\hline
\end{tabular}
\caption{Parameters used in the calculation.}
\label{parameter}
\end{table}

\subsection{QD's in  (001) quantum well}

\subsubsection{Small well width without strain}

We first consider QD's in a small (001)  quantum well ($a=5$\ nm)
without any strain where the lowest eigen states of $H_0$ are heavy
holes and the separation between the heave and light holes is
around 0.1\ eV. Due to the small well width, only the lowest
subband is needed in the calculation.  
Therefore, as pointed out above,
only the hole-phonon scattering due to the deformation potential 
contributes to the spin relaxation.

\begin{figure} 
\centering
  \psfig{figure=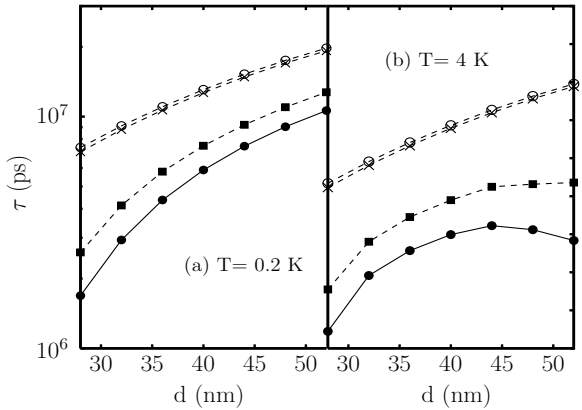,width=8cm,height=5.5cm,angle=0}
  \label{1}
  \caption{SRT {\em vs.} QD diameter $d$. Curve with { $\bullet$}: exact
    diagonalization result with the energy sufficiently converged;
    Curve with { $\times$}: Perturbation result; Curve
    with {  $\bigcirc$ }: exact diagonalization result but with
    only the lowest two heavy hole and 
the lowest two light hole levels used as basis functions. Curve
    with  {\tiny $\blacksquare$}: exact  diagonalization result but with
  only the 16 energy levels of $H_0$ given in the text as basis functions. 
(a): $T=0.2$ K and (b): $T=4$ K 
    .}
\end{figure}

In Fig.\ 1 we plot the  SRT as a function of the QD diameter
for two temperatures.
 $B=1$\ T in the calculation.
Curves with {\LARGE $\bullet$} are the results
obtained by the exact diagonalization method with the energy levels
sufficiently converged. Unexpectedly, differing from intuition as
well as the results of an electron spin in QD's,\cite{Cheng}
it shows in Fig.\ 1(a) that the SRT
{\em increases} with the QD diameter. 

To understand/check this result,
we compare the results from the exact diagonalization method
with those from the perturbation approach widely used in the 
literature,\cite{Woods,Khaetsii,Alexander,Alexander1} but
with the proper modification by including the second order
corrections to the energy spectrum as pointed out in our
previous work.\cite{Cheng} 
Here we treat the off-diagonal part
of Luttinger Hamiltonian $H_h$ as perturbation and
calculate the SRT between the lowest
two energy levels composed by two lowest heavy hole and
two lowest light hole states of $H_0$:
$|0,0,1,\sigma\rangle$ with $\sigma = \pm \frac{3}{2}$ and 
$\pm\frac{1}{2}$.
The wave functions can be written as:
\begin{eqnarray}
  \label{wave_lowest_2}
  \langle{\bf r}|\Psi_{\uparrow}\rangle
&=&\langle{\bf r}|0,0,1,\frac{3}{2}\rangle-{\cal A}\langle
{\bf  r}|0,0,1,-\frac{1}{2}\rangle\nonumber\ ,\\
\langle{\bf r}|\Psi_{\downarrow}\rangle &=& 
\langle{\bf r}|0,0,1,-\frac{3}{2}\rangle -{\cal
    B}\langle {\bf r}|0,0,1,\frac{1}{2}\rangle \ ,
\end{eqnarray}
in which 
  \begin{eqnarray}
\label{A}
    {\cal A}=-
    \frac{\sqrt{3}\gamma_3\hbar
    eB}{4\gamma_2 \hbar^2 (\pi^2/a^2-\alpha)-4\hbar e B \kappa}\ ,\\ 
\label{B}
    {\cal B}=-
    \frac{\sqrt{3}\gamma_3\hbar
      eB}{4\gamma_2 \hbar^2 (\pi^2/a^2-\alpha)+4\hbar e B \kappa}\ .
  \end{eqnarray}
With the second order correction to the energy included, the energy
difference between the spin-up $|\Psi_\uparrow\rangle$ 
 and spin-down $|\Psi_\downarrow\rangle$ states  can be
written as: 
\begin{eqnarray}
  \label{DeltaE}
  \Delta E&=&\frac{6\hbar eB\kappa}{2m_0}+|{\cal
    A}|^2(E_{0,0,1,\frac{3}{2}}-E_{0,0,1,\frac{1}{2}})\nonumber\\ && - |{\cal
    B}|^2(E_{0,0,1,-\frac{3}{2}}-E_{0,0,1,-\frac{1}{2}})\nonumber\\
    &=& \frac{6\hbar eB\kappa}{2m_0}-
    \frac{3\kappa\gamma_3^2 (\hbar e
    B)^3/(4m_0)} {[\gamma_2\hbar^2
    (\pi^2/a^2-\alpha)]^2 - (\hbar e B \kappa)^2}\ ,
\end{eqnarray} 
in which the first term represents Zeeman splitting.
The SRT $\tau$ can therefore be written as:
\begin{eqnarray}
    \label{tau}
    \frac{1}{\tau}
    &=&\sum_{\lambda}\frac{D_d^2 \Delta E^3{\bar {n_{Q\lambda}}}}{2\pi(\hbar
    v_{s\lambda})^4 v_{s\lambda}\rho} 
    \int^{\frac{\pi}{2}}_{0}d\theta 
{\cal K}_{\lambda}{(\theta)}
 ({\cal A}-{\cal B})^2
\nonumber\\ &&\mbox{}\times \exp(-\frac {q^2}{ 2 \alpha}) I^2(q_z)\ ,
  \end{eqnarray}
with $Q=\Delta E/(\hbar v_{s})$, $q=Q\sin{\theta}$, $q_z=Q\cos{\theta}$ and
$I(q_z)={8\pi^2\sin(aq_z/{2})}/\{aq_z[4\pi^2-(aq_z)^2]\}$.
$\lambda$ stands for the branch of phonon: for
longitude mode ${\cal K}_{l}{(\theta)}=\sin^3(\theta)
\cos^2(\theta)$ and for two
transverse modes ${\cal K}_{t1}{(\theta)}=\frac{1}{4}
\sin(\theta)\cos^2(2\theta)$ and
${\cal K}_{t2}{(\theta)}=\frac{1}{4}\sin(\theta)
\cos^2(\theta)$ respectively. The SRT's
calculated from Eq.\ (\ref{tau}) are plotted as the curve with
{$\times$} in Fig.\ 1(a) which coincides with the curve
with {\Large $\circ$}  obtained from the exact 
diagonalization method but with exactly the same four basis functions
used in the perturbation method as basis.
One can see that the SRT $\tau$ does {\em increase} with the QD diameter at 
$T=0.2$ K. From Eqs.\ (\ref{A}-\ref{tau}) one finds that the
SRT depends on the diameter only through
 $\alpha$, which can be approximated into
 $\alpha=1/d^2$ in the case $\omega_0\gg\omega_B$. Moreover, the
mixture of wave functions and the energy
difference $\Delta E$ hardly change  with $d$ for $\pi^2/a^2\gg
1/d^2$. Therefore only the exponential term $\exp(-\frac {q^2}{ 2 \alpha})$ in
Eq.\ (\ref{tau}) decreases with $d$. As a result larger QD diameter
corresponds to longer SRT at low temperature.

As shown in our previous work that the right perturbation approach
(with the second order corrections to energy spectrum included)
with the lowest few levels of $H_0$ as basis functions
may lead to totally opposite trend from the
exact diagonalization method with sufficient
number of basis functions.\cite{Cheng} In order  to rule out this
possibility in the present analysis,  we plot in Fig.\ 1(a)
the SRT calculated from exact diagonalization method
but with sixteen eigen functions of $H_0$ as basis functions,
{\em i.e.}, $|0,0,1,\sigma\rangle$ and $|0,1,1,\sigma\rangle$
with $\sigma=\pm \frac{1}{2}$ and $\pm \frac{3}{2}$.
It is seen from the figure that it produces the same $\tau$-$d$
dependence.

In Fig.\ 1(b) we plot the same curves but at $T=4$ K. It is seen
that at this temperature the SRT has a maximum 
as a function of  the diameter $d$. This is because at high temperatures
the scattering between the higher energy levels becomes important.
These high energy levels are arrayed very close to each other.
The increase of diameters makes more levels into the
scattering channels and thus induces a faster spin relaxation
if it overcomes the opposite tendency described above.
The perturbation approach only includes  the
lowest two heavy-hole and two light-hole  energy levels
and therefore can not get the maximum feature here.

\begin{figure}[htb]
\centering
  \psfig{figure=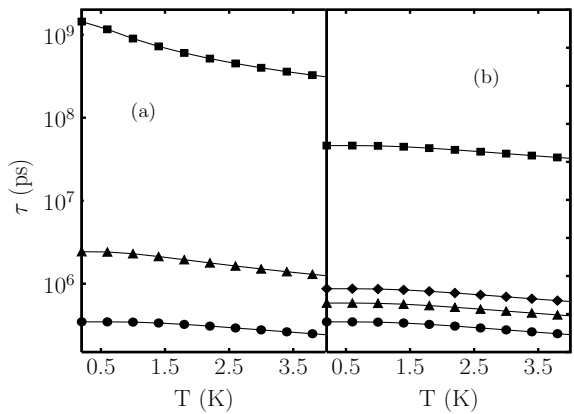,width=8cm,height=5.5cm,angle=0}
  \caption{SRT {\em vs.} temperature $T$ for QD
with  $a=5$ nm and $d=20$ nm. (a):
 Results under different magnetic fields. Curve
  with {\tiny $\blacksquare$}: $B=0.2$ T; Curve with $\blacktriangle$:
   $B=0.6$ T; Curve with {\large $\bullet$}: $B=1$ T. (b): 
Results at  $B=1$ T with contributions from different branches of phonons 
  specified. Curve
  with {\tiny $\blacksquare$}: Contribution from longitudinal 
 phonons;  Curve with $\blacktriangle$: Contribution from
  the transverse  phonons of first 
branch;  Curve with $\blacklozenge$: Contributions from
  the transverse phonons of the second 
branch; Curve with {\large $\bullet$}: The total SRT.}
  
\end{figure}

In Fig.\ 2(a) we plot the SRT as a function of the temperature for a
QD with $a=5$ nm and $d=20$ nm under three different magnetic
fields. From the figure one finds that the SRT decreases with the
temperature as  higher temperature leads to larger number of phonons
 $n_{Q\lambda}$ and consequently a larger transition probability. 
Moreover, as pointed out in our previous work,\cite{Cheng} smaller magnetic
field makes the SRT decrease faster with the temperature.
This is because the energy intervals between different energy levels are small
in the presence of a small magnetic field. This 
 leads to a faster response to the temperature. In Fig.\ 2(b) we
further specify the contributions from different branches of phonons.
It is stressed again that only the hole-phonon scattering due to
the deformation potential contributes to the spin relaxation here. As
the temperature is below 4 K, the scattering from spin-up to 
spin-down states, in which phonons are emitted, is much larger 
than that from the reverse process, 
in which phonons are absorbed. Therefore, unless specified,
the scattering rate $\frac{1}{\tau}$ is defined to be the scattering
from spin-up to spin-down states throughout this paper. 
It is seen from the figure that the SRT here is determined by the
transverse modes.

\subsubsection{Small width with strain} 

\begin{figure}[htb]
  \psfig{figure=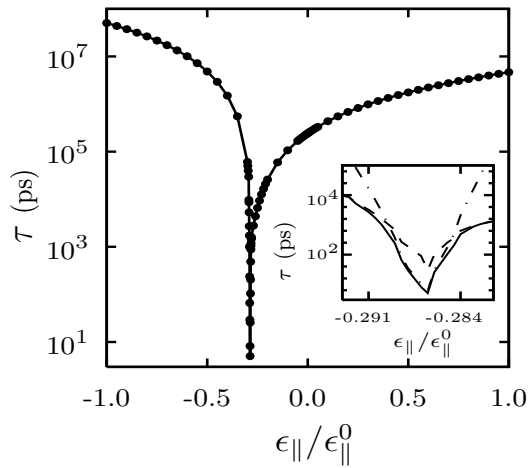,width=7.cm,height=6.2cm,angle=0}
  \caption{SRT versus strain for a QD in (001) quantum well
 at  $a=5$\ nm, $d=20$\ nm and $B=1$\ T at $T=4$ K. 
Solid curve: total SRT; Chain curve: 
SRT induced by the hole-phonon scattering
due to  piezoelectric coupling; Dashed curve: 
SRT induced by the hole-phonon scattering
due to deformation potential.}
  
\end{figure}

As pointed out
in the previous section, strain on (001) quantum well can change
or even reverse the relative  
positions of energy levels of heavy and light holes.
Now we turn to  investigate QD's under different strains in (001) quantum well
with $a=5$\ nm, $d=20$\ nm and $B=1$\ T at $T=4$\ K.   The strain
is adjusted by changing the strain tensor
component $\epsilon_{\|}$ in Eqs.\ (\ref{001})
and (\ref{001z}). We plot the SRT versus $\epsilon_{\|}/\epsilon_{\|}^0$ 
in Fig.\ 3 with 
 $\epsilon_{\|}^0$ obtained by substituting the
lattice constants of GaAs and InAs
for $a_2$ and $a_1$ respectively  in Eq.\ (\ref{001}).
It is seen from the figure that when $\epsilon_{\|}>0$,
the SRT $\tau$ increases with applied strain.
This is because the positive strain enhances the gap between
the heavy hole and light hole. Nevertheless, the flip
from the spin-up heavy hole to the spin-down
one is determined by the light-hole components in the
wavefunctions. Increasing the
gap between the heavy and light holes greatly reduces 
the spin relaxation and leads to the increase of SRT.
When we apply a negative strain, as the gap decreases, the SRT
decreases as shown in the figure. Particularly at 
 $\epsilon_{\|}/\epsilon_{\|}^0=-0.3$ the lowest two energy
states change from the heavy-hole states to the light-hole ones
and the SRT shows  a minimum. When  $\epsilon_{\|}/\epsilon_{\|}^0<-0.3$
the initial spin states are light holes and increasing strain along the
negative direction enhances the gap of light hole and heavy hole again and
therefore the SRT is enhanced again. 

SRT around  $\epsilon_{\|}/\epsilon_{\|}^0=-0.3$ needs more address.
Around this point, the energy levels of heavy hole and light hole are
close to each other and therefore as said before that 
the hole-phonon scattering due to piezoelectric coupling is
able to contribute to the spin relaxation. This can be seen in the 
inset of Fig.\ 3 where SRT due to the piezoelectric coupling is
plotted as a chain curve and that due to deformation
potential is plotted as dashed one. The solid curve is the total 
SRT.  Very close to $\epsilon_{\|}/\epsilon_{\|}^0=-0.3$, the SRT is
determined by the hole-phonon scattering due to piezoelectric coupling.
However, the contribution of piezoelectric coupling decays
dramatically with little deviation  of the strain from $-0.3$.
When  $\epsilon_{\|}/\epsilon_{\|}^0<-0.29$ or $>-0.284$, the SRT
is totally determined by the hole-phonon scattering 
due to the deformation potential.

\subsubsection{Large width}

\begin{figure}[htb]
  \psfig{figure=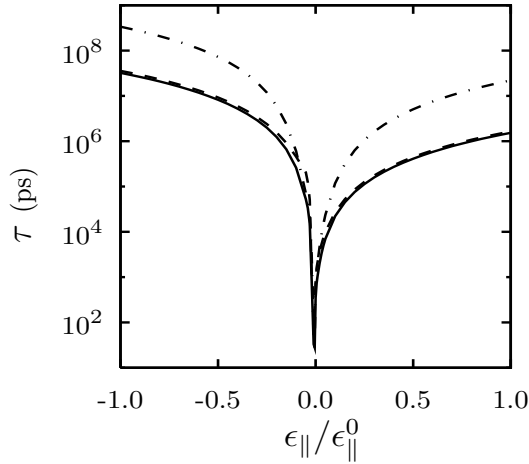,width=7.cm,height=6.2cm,angle=0}
  \caption{SRT {\em vs.}  strain  for a QD in (001) quantum well
    at  $a=20$ nm, $d=20$ nm and $B=1$ T at $T=4$ K. 
    Solid curve: total SRT; Chain curve: 
    SRT induced by the hole-phonon scattering
    due to  piezoelectric coupling; Dashed curve: 
    SRT induced by the hole-phonon scattering
    due to deformation potential.}
  
\end{figure}

Now we turn to  investigate the SRT in a QD of $a=20$ nm,
$d=20$\ nm and $B=1$\ T at $T=4$\ K.
For such a well width, one has to include states with
$n_z\ge 2$. From Eqs.\ (\ref {Luttinger_Hamiltonian}-\ref{S}) 
one can see that the scattering between
different subbands makes  $\pm\frac{3}{2}$ states mix 
with  $\pm\frac{1}{2}$ states.
Therefore, hole-phonon scattering due to the
piezoelectric coupling makes
contribution to the spin-flip scattering 
with or without strain.

In Fig.\ 4, we plot the SRT as
function of  $\epsilon_{\|}/\epsilon_{\|}^0$.
It is seen from the figure that the SRT shows a minimum around
$\epsilon_{\|}/\epsilon_{\|}^0\sim -0.01$. This is because 
for large quantum well, the lift of the $\Gamma$-point
degeneracy is very small and the lowest heavy-hole and light-hole
states are very close to each other.
Again, the SRT increases with the applied positive/negative strain 
due to the separation of the heavy hole and light hole states.
Differing from the case of single subband, here the hole-phonon
scattering due to the piezoelectric coupling 
makes strong contribution to the SRT even without strain.
Nonetheless it is shown in the figure that when the
strain $\epsilon_{\|}/\epsilon_{\|}^0$ 
is slightly deviated from $-0.01$ the SRT is mainly determined 
by the hole-phonon scattering due to deformation potential.

\begin{figure}[htb]
\centering
  \psfig{figure=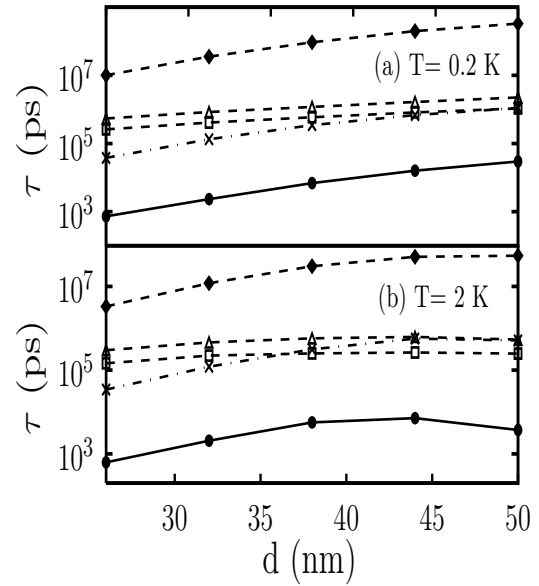,width=7.cm,height=8.5cm,angle=0}
  \caption{SRT {\em vs.} the diameter $d$ for QD's
    at $a=20$ nm and $B=1$ T under different 
    strains. Curve with {$\bullet$}: $\epsilon_{\|}/\epsilon_{\|}^0=0$; 
    {$\times$}: $\epsilon_{\|}/\epsilon_{\|}^0=0.07$; {$\square$}:
    $\epsilon_{\|}/\epsilon_{\|}^0=-0.07$;  
{$\vartriangle$}: $\epsilon_{\|}/\epsilon_{\|}^0=-0.09$;
    {$\blacklozenge$}:$\epsilon_{\|}/\epsilon_{\|}^0=-0.3$. (a) $T=0.2$
    K; (b) $T=2$ K. }
 
\end{figure}

In Fig.\ 5 we plot the SRT as a function of QD diameter $d$ when
$a=20$\ nm under different strains with solid curves for 
the strain-free case, chain curves for positive strain cases and
dashed curves for negative strain cases. Similar to the case of
small well width without strain in Fig.\ 1, 
for large well width without strain
here  the SRT also increases with the diameter 
for low temperature ($T=0.2$ K) [Fig.\ 5(a)] and 
shows a maximum for higher temperature ($T=2$\ K) [Fig.\ 5(b)].
Strains keep these trends. Nevertheless for small negative strain 
$-0.2<\epsilon_{\|}/\epsilon_{\|}^0 < 0$ 
the variations become smoother.
 This feature can be understood as follows:
For negative strain the heavy hole states intercept
with the light hole ones at high energies. 
The states around these intercepting points are particular
efficient in spin relaxation.  As said in
the previous section, the high energy states are arrayed very close to
each other. The increase of diameter drives more states into the
scattering channel and partly compensates the tendency of decrease.
 For larger negative strain ($\epsilon_{\|}/\epsilon_{\|}<-0.2$),
the heavy and light holes are separated again and the change of the
SRT with the diameter becomes fast again as shown in the figure.

\begin{figure}[htb]
  \psfig{figure=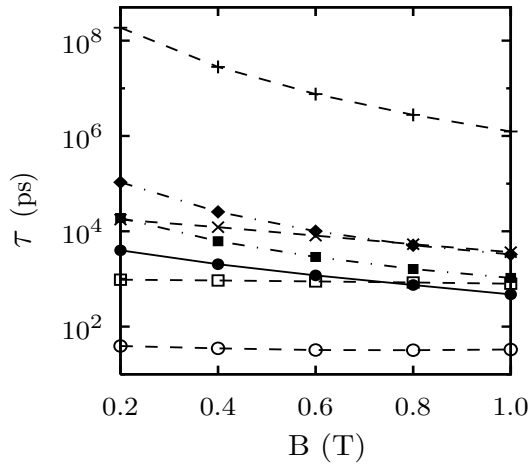,width=7.cm,height=6.2cm,angle=0}
  \caption{SRT {\em vs.} the magnetic field $B$ for a QD in (001)
    quantum well  at $a=20$ nm, $d=20$ T and $T=4$ K under different 
    strains. Curve with {$\bullet$}: 
    $\epsilon_{\|}/\epsilon_{\|}^0=0$; {\tiny
      $\blacksquare$}: $\epsilon_{\|}/\epsilon_{\|}^0=0.01$; 
    {$\blacklozenge$}:
    $\epsilon_{\|}/\epsilon_{\|}^0=0.03$; 
    {$\bigcirc$}:
    $\epsilon_{\|}/\epsilon_{\|}^0=-0.01$; {$\square$}: 
    $\epsilon_{\|}/\epsilon_{\|}^0=-0.02$; {$\times$}: 
    $\epsilon_{\|}/\epsilon_{\|}^0=-0.03$;
    $+$: $\epsilon_{\|}/\epsilon_{\|}^0=-0.2$.}
  
\end{figure}

The SRT as a function of magnetic field $B$ under different
strains at $T=4$ K is plotted in Fig.\ 6. 
As shown in Eqs.\ (\ref{A}) and (\ref{B}),
the spin mixing is enhanced with the
increase of  the magnetic field. Therefore $\tau$ decreases with $B$.
Moreover, similar to the 
 case of the diameter dependence of the SRT, one finds
that the $\tau$-$B$ dependence becomes weak 
when a negative strain is in the range
$-0.2<\epsilon_{\|}/\epsilon_{\|}^0 < 0$.
This is because when the heavy-hole states intercept with the
 light-hole ones,  Zeeman
splitting which appears in the denominators of the
spin mixing coefficients [{\em e.g.}, in
the  denominators of Eqs.\ (\ref{A}) and (\ref{B})]
becomes important which partially compensates the increase of the
spin mixing with the magnetic field ({\em ie.},  $B$ in numerators
of spin mixing coefficients).
 Therefore, the SRT changes with $B$ slowly.
However, for  a larger negative
strain which makes a big separation  between the heavy hole 
and light hole states, the $\tau$-$B$ dependence becomes stronger
again.

\subsection{QD's in (111) quantum well}

\begin{figure}[htb]
  \psfig{figure=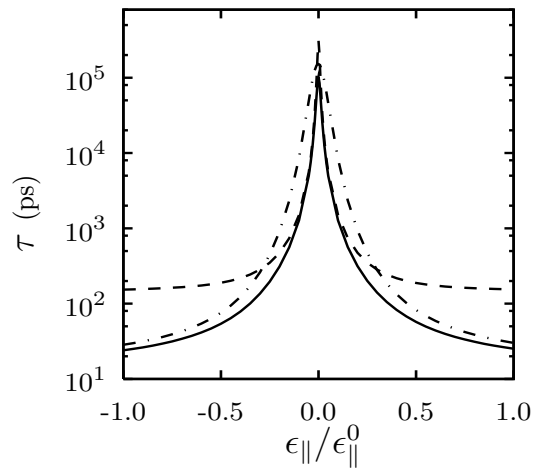,width=7cm,height=6.2cm,angle=0}
  \caption{SRT {\em vs.} strain at  $a=5$ nm, $d=20$ nm, $B=1$ T and $T=4$ K.
     Solid curve: total SRT; Chain curve: 
    SRT induced by the hole-phonon scattering
    due to  piezoelectric coupling; Dashed curve: 
    SRT induced by the hole-phonon scattering
    due to deformation potential.}
  
\end{figure}

We now turn to investigate QD's under different strains in (111) quantum well.
As pointed out before, differing from the case of (001) quantum wells,
there always exists spin mixing between spin-up
and -down heavy (light) holes in the eigen functions of the 
Luttinger Hamiltonian of (111)-oriented quantum wells.
Therefore, hole-phonon scattering due to both deformation
potential and 
piezoelectric coupling contributes to the spin relaxation even there
is no strain on QD. Moreover, the effect of strain on (111)-oriented crystal is
also different from the (001) case: The strain hardly changes the relative
position of energy levels of heavy and light holes, but introduces
additional spin mixing which leads to additional spin relaxation.

In Fig.\ 7 we plot the SRT versus  $\epsilon_{\|}/\epsilon_{\|}^0$
in a QD of $a=5$\ nm, $d=20$\ nm and $B=1$\ T at $T=4$\ K.  
It is seen from the figure that opposite to the (001) case as shown
 in Figs.\ 3 and 4, here the SRT {\em decreases} rapidly 
with the increase of strain in both positive and negative directions. 
This is because that the 
additional spin mixing introduced by the strain is the main effect
in the present case, which makes the scattering rate increases with strain.
Moreover, one finds that for small strain, the SRT is determined by the 
hole-phonon scattering due to deformation potential but after 
$|\epsilon_{\|}/\epsilon_{\|}^0|>0.1$, 
hole-phonon scattering due to the piezoelectric coupling starts to
contribute to the spin relaxation and after 
$|\epsilon_{\|}/\epsilon_{\|}^0|>0.3$ it takes over the
scattering due to the deformation coupling and becomes the
leading contribution. However  both contributions should be
included in the calculation when strain is presented in (111) quantum
wells.

\begin{figure}[hbt]
\centering
  \psfig{figure=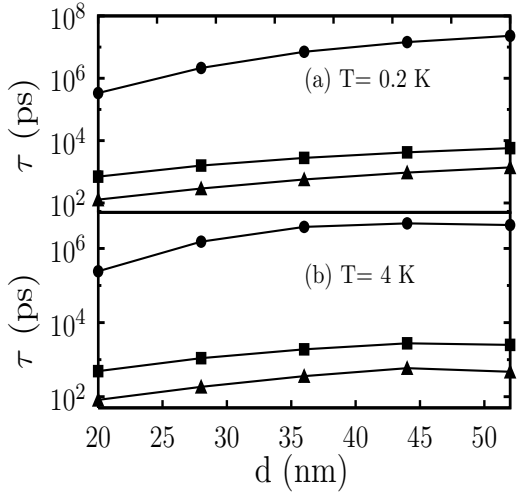,width=7.cm,height=7.cm,angle=0}
  \caption{SRT {\em vs.} the diameter $d$ for a QD in (111) quantum well
    at $a=5$ nm and $B=1$ T under different
    strains.  Curve with {$\bullet$}:
    $\epsilon_{\|}/\epsilon_{\|}^0=0$; {\tiny $\blacksquare$}:
    $\epsilon_{\|}/\epsilon_{\|}^0=0.12$; 
    {$\blacktriangle$}: $\epsilon_{\|}/\epsilon_{\|}^0=0.28$. (a):
    $T=0.2$ K, (b) $T=4$ K.}
\end{figure}

We discuss the diameter and magnetic field dependence of SRT under
different strains.  In Fig.\ 8 the SRT is plotted against
the QD diameter $d$ for different strains.
As the SRT with positive and negative strains
are symmetrical, we only show the case
with $\epsilon_{\|}/\epsilon_{\|}^0 \ge 0$.
It is noted that when $\epsilon_{\|}/\epsilon_{\|}^0=0.12$ the 
hole-phonon scattering due to the deformation potential
is  dominant but when it is  $0.28$  the scattering due to the
piezoelectric coupling
becomes more important. Similar to the cases in the previous
sections, the SRT increases with diameter monotonously when $T=0.2$ K
[Fig.\ 8(a)] and has a maximum when $T=4$ K [Fig.\ 8(b)].

\begin{figure}[htb]
  \psfig{figure=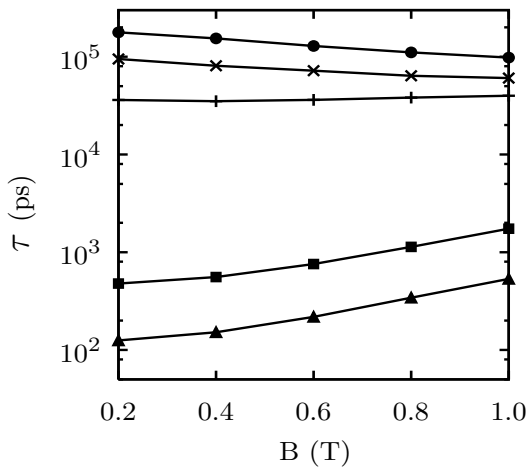,width=7cm,height=6.2cm,angle=0}
  \caption{SRT {\em vs.} the magnetic field $B$
    at $a=5$ nm, $d=20$ T and $T=4$ K under different
    strains. Curve with {$\bullet$}:
    $\epsilon_{\|}/\epsilon_{\|}^0=0$; {$\times$}:
    $\epsilon_{\|}/\epsilon_{\|}^0=0.01$; {$+$}:
    $\epsilon_{\|}/\epsilon_{\|}^0=0.02$; {\tiny $\blacksquare$}:
    $\epsilon_{\|}/\epsilon_{\|}^0=0.12$; 
{$\blacktriangle$}: $\epsilon_{\|}/\epsilon_{\|}^0=0.28$.}  
\end{figure}

In Fig.\ 9 we show the magnetic field dependence of the SRT
under different strains. When there is no strain or very small
strain, similar to the case of QD in (001) quantum well and our
previous investigation on electron spin in QD's,\cite{Cheng}
 the SRT decreases with the magnetic field.
However for a little bigger strain the  SRT {\em increases} with the 
magnetic field. This is understood that the spin mixing
induced by strain in (111) quantum well makes a major contribution
to the spin relaxation for sufficient big strain. Nevertheless,
this mixing  {\em decreases} with $B$. This can be seen in following:
In the case when there is no inter subband spin mixing and the
heavy- and light-hole states are separated from each other, then
almost all the spin mixing comes from the 
off-diagonal terms of the strain  Hamiltonian [Eqs.\ (\ref{H})
and (\ref{I})]. Using the perturbation method, and adopting 
the lowest four states of $H_0$ as  basis,
the wave functions are written into:
\begin{eqnarray}
  \label{111_wave_lowest_2}
  \langle{\bf r} |\Psi_{\uparrow}
\rangle&=&\langle {\bf r}|0,0,1,\frac{3}{2}\rangle+{\cal C}\langle
  {\bf r}|0,0,1,\frac{1}{2}\rangle+{\cal D}\langle
  {\bf r}|0,0,1,-\frac{1}{2}\rangle\nonumber\ ,\\
\langle{\bf r} |\Psi_{\downarrow}\rangle
 &=& \langle {\bf r}|0,0,1,-\frac{3}{2}\rangle +{\cal
    E}\langle {\bf r}|0,0,1,\frac{1}{2}\rangle\nonumber\\&&+{\cal F}\langle
  {\bf r}|0,0,1,-\frac{1}{2}\rangle \ ,
\end{eqnarray}
in which ${\cal C}=m_0 H/(\hbar e B \kappa+\Delta\epsilon)$, ${\cal
  D}=m_0 I/(2\hbar e B \kappa+\Delta\epsilon)$, ${\cal E}=-m_0
  I^{\ast}/(2\hbar e B \kappa+\Delta\epsilon)$  and
  ${\cal F}=m_0 H^{\ast}/(\hbar e B \kappa+\Delta\epsilon)$ with
  $\Delta\epsilon= 2\gamma_2\hbar\Omega
-4\gamma_2\frac{\hbar^2\pi^2}{m_0a^2}$. $H$ and $I$ 
are the matrix elements of strain Hamiltonian [Eqs.\ (\ref{H})
and (\ref{I})], which are independent of the magnetic field.
Consequently the spin mixing decreases with $B$.

\begin{figure}[htb]
  \psfig{figure=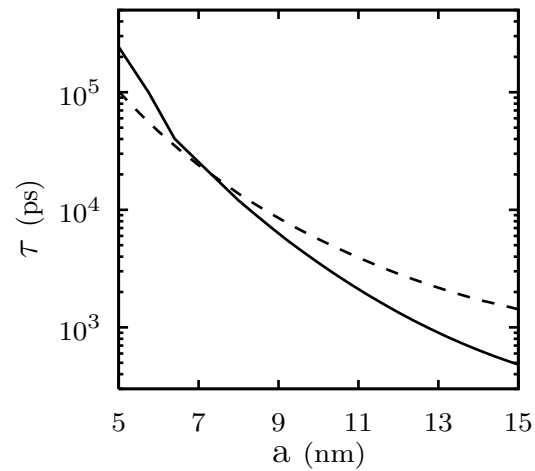,width=7.cm,height=6.2cm,angle=0}
  \caption{SRT versus the well width $a$
    at $d=20$ nm, $B=1$ T and $T=4$ K. 
Solid curve:  (001) quantum well; Dashed curve:  (111)
      quantum well.} 
  
\end{figure}
  
\subsection{Well width dependence of the SRT}

Finally we investigate the quantum well width dependence of the
SRT of QD's with $d=20$ nm and $B=1$ T at $T=4$ K. 
The SRT's of QD's in (001) and (111)
quantum wells are plotted
as function of quantum well width $a$.
It is seen that for both cases the SRT's decrease with the
well width, which is totally opposite to the 
cases of electron spin in QD's\cite{Cheng} and quantum wells.\cite{wu}
This difference originates from the fact that for electron
spin the spin-orbit coupling decreases dramatically 
with the well width.\cite{Cheng,wu} Nevertheless, for hole spin
although the spin-orbit coupling also decreases with $a$
[see Eqs.\ (\ref{S}) and (\ref{111S})], the decrease of the
intervals between different energy states is faster [see Eqs.\ (\ref{PQ})
and (\ref{111PQ})]. Consequently more states are included in the
spin-flip scattering channel and $\tau$ decreases with the well width.

\section{Conclusions}
In conclusion, we have performed a comprehensive investigation
on  hole spin relaxation in GaAs QD's
confined in  quantum wells along (001) and (111) directions
by exactly diagonalizing the hole
Luttinger Hamiltonian.

We find for QD's in (001) quantum wells with small well width
where only the lowest subband is involved, 
the SRT {\em increases} with the QD diameter at very low temperature
({\em e.g.}, 0.2\ K) or first increases until it reaches a maximum and then
decreases at higher temperature ({\em e.g.}, 4\ K).
These features are opposite to those of electron spin in QD's.
Moreover, unlike the case of electron spin where
the SRT is mainly determined by the electron-phonon scattering 
due to the piezoelectric coupling, 
here only the hole-phonon scattering due to the deformation
potential contributes to the spin relaxation. 
Strain changes the relative positions of energy levels of heavy hole
and light hole. A positive strain increases the energy gap between
heavy hole and light hole and enhances the SRT. A negative strain
decreases the gap and reduces the SRT until the interchange of
the lowest energy states from heavy hole to light hole. After that
the SRT increases again. Moreover, very close to the interchange point, as
the energy levels of heavy hole and light hole 
are very close to each other, the
hole-phonon scattering due to piezoelectric coupling contributes to
the spin relaxation too. For large well width where multi-subband effect
is important, hole-phonon scattering
due to the piezoelectric coupling contributes to the spin-flip
scattering with or without strain. Nevertheless
the SRT is still mainly determined by the scattering due to
deformation potential except at the interchange point.
The magnetic field dependence of the SRT is also discussed.

For QD's in (111) quantum well things are quite different from
those for QD's in  (001) quantum well: Hole-phonon scattering
due to both piezoelectric coupling and deformation potential
contributes to the spin relaxation and should be both included in the
calculation, regardless of the well width; Strains can hardly change 
the relative positions of energy levels of heavy hole and light hole but
introduce {\em additional} spin mixing. Therefore
the SRT decreases rapidly with strain; The SRT 
decreases with magnetic field like the case of QD in (001) quantum well and our
previous investigation on electron spin in QD's when there is
no strain or very small strain.  However for 
strain which is big enough that the spin mixing is mainly determined by
it, the SRT {\em increases} with $B$.

Finally we show that the hole SRT
decreases with well width for QD's in both (001) and (111) 
quantum wells  which is totally opposite to the 
cases of electron spin in QD's and quantum wells. 

\acknowledgments

This work was supported by the Natural Science Foundation of China
under Grant No. 90303012.
MWW was also supported by the ``100 Person Project"
of Chinese Academy of Sciences and the Natural Science Foundation of China
under Grant No. 10247002. He would like to thank B. F. Zhu for valuable
discussions.

\end{document}